\documentclass{article}
\usepackage{dafne99pro}
\begin{document}
\title{ 
FUTURE PROSPECTS FOR CP VIOLATION IN HADRON MACHINES
}
\author{
Ramon Miquel \\
{\em Departament d'Estructura i Constituents de la Mat\`eria and IFAE} \\
{\em Facultat de F\'{\i}sica, Universitat de Barcelona} \\ 
{\em Diagonal, 647, E-08028 Barcelona, Spain}
}
\maketitle
\baselineskip=11.6pt
\begin{abstract}
The capabilities of future CP violation experiments in hadron machines 
are reviewed. Special emphasis is put on the CDF experiment in the Run II of 
the Tevatron, due to start taking data in 2001,
and on LHCb and BTeV, on a longer time scale.
\end{abstract}
\baselineskip=14pt
\section{Introduction}
While the first statistically significant observation of CP violation in the B
meson system will most likely take place in one of the $e^+e^-$ $b$-factory
experiments (BaBar, Belle), this talk will try to prove that the contribution
from experiments in hadron colliders will be dominant in the future, starting
in 2001.

Figure~1 shows the standard CP violation triangle. The angle $\beta$ is the 
most easily accessible, in particular through the asymmetry in the 
decays $B^0/\overline{B}^0 \to J/\Psi K_S$.
First determinations by CDF~\cite{cdf-2b98}, 
OPAL~\cite{opal-2b} and ALEPH~\cite{aleph-2b} have
recently become available. When combined, they give a two-standard deviation
measurement of $\sin 2\beta$:
\begin{equation}
\sin 2\beta = 0.78 \pm 0.37\ .
\end{equation}
\begin{figure}[t]
 \vspace{7.0cm}
\includegraphics{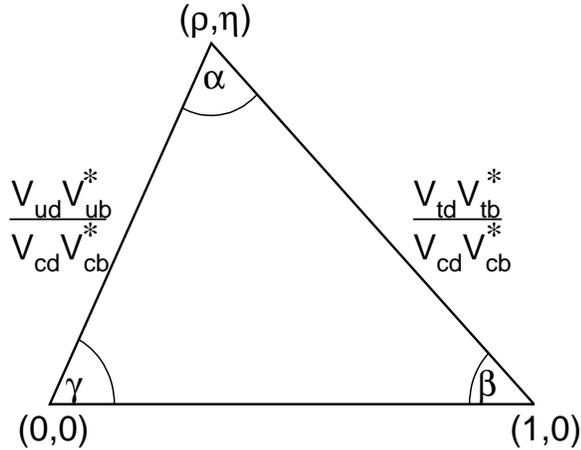}
 \caption{\it
      The standard CP violation triangle.
    \label{fig1} }
\end{figure}
The combination is totally dominated by the CDF result. By summer 2000,
the electro-positron 
colliders should get a measurement with
an uncertainty about $\pm 0.15$~\cite{babar-d99}.
The task is substantially more involved for the other two angles, $\alpha$ 
and $\gamma$. It is not obvious when will BaBar or Belle be able to
obtain a measurement.

Hadronic uncertainties diminish the usefulness of a 
determination of the length of the side of the triangle opposite $\beta$.
The length of the side opposite $\gamma$, can be obtained from the
determination of $\Delta m_s$, the frequency of the oscillations in the 
$B^0_s-\overline{B}^0_s$ system. The current limit from LEP is~\cite{lep-xs}:
\begin{equation}
x_s = \frac{\Delta m_s}{\Gamma_s} > 21.8      \ .
\end{equation}
Standard Model expectations for $x_s$ are in the range 20--30~\cite{xs-sm}. 
When the determination of $x_s$ will become available, it will be possible to
combine it with $x_d$, the similar parameter in the $B_d$ system, now
measured to be~\cite{lep-xs}
\begin{equation}
x_d = \frac{\Delta m_d}{\Gamma_d} = 0.717 \pm 0.026       \ ,
\end{equation}
in order to obtain the length of the side of the triangle opposite $\gamma$
through the relation:
\begin{equation}
\frac{\Delta m_d}{\Delta m_s} 
\simeq \frac{\left|V_{td}\right|}{\left|V_{ts}\right|} 
\simeq \frac{\left|V_{td}\right|}{\left|V_{cb}\right|} \ .
\end{equation}
Since the lepton machines are only scheduled to run at the energy of
the $\Upsilon$(4S) resonance, no $B_s$ mesons will be produced. Therefore, 
$x_s$ will only be determined in hadron machine experiments.

The standard CP triangle in fig.~1 is the only one obtained from the standard
Wolfenstein parametrization of the CKM matrix~\cite{ckm-matrix}. However, one
should remember that this is just an expansion of the complete matrix in powers
of $\lambda$, the sinus of the Cabibbo angle, valid up to 
order ${\cal O}(\lambda^3)$. If one proceeds with the expansion 
up to $\lambda^4$, one encounters a new imaginary contribution to $V_{ts}$, 
which gives raise to a new unitarity triangle.
%shown in fig.~2.
%%
%\begin{figure}[t]
% \vspace{9.0cm}
%\special{psfile=dafne99profig.ps voffset=-220 hoffset=-30
%         hscale=80 vscale=80 angle=0}
% \caption{\it
%      The new CP violation triangle.
%    \label{fig2} }
%\end{figure}
%%
A new angle, $\delta\gamma$, appears, which
in the Standard Model is proportional 
to $\lambda^2$, and, therefore, very small, of order ${\cal O}(10^{-2})$. It is
easily accessible through the decay $B_s \to J/\Psi \phi$, the $B_s$
equivalent to the ``golden channel'' $B_d \to J/\Psi K_S$. Again, this is not
possible in the lepton machines, which do not produce $B_s$.

The experimental program in the coming years has two clear goals:
\begin{itemize}
\item First, to observe for the first time CP violation in the B system. As 
explained above, this will most likely be achieved by BaBar and/or Belle,
followed by HERA-B at DESY, CDF and D0.
\item Then, to
test whether CP violation is generated by the Standard Model. Here,
all the experiments mentioned before will contribute but, most probably, the
second-generation dedicated experiments, LHCb and BTeV, will be needed.
\end{itemize}

In the remaining of the talk, the capabilities of the experiments foreseen
in future hadron colliders will be reviewed. Section~2 will cover the near 
future: CDF and D0 at the Fermilab Tevatron,
scheduled to start physics data-taking in spring
2001. The more distant future, from 2005 on, will be covered in Section~3.
ATLAS, CMS and LHCb are approved to take data at CERN's LHC, while BTeV, if
approved, will run at the Tevatron. Special emphasis will be put in
the two dedicated $b$-experiments, LHCb and BTeV. Finally, section~4 will give
a short summary of the talk.
\section{The Near Future: CDF and D0}
\subsection{The detector upgrades}
The upgraded Tevatron at Fermilab is scheduled to start collider
physics operation in
spring 2001 with two upgraded detectors, CDF and D0. The luminosity delivered
in the first two years of operation is expected to reach about 2~fb$^{-1}$, 
which is equivalent to a production of about 10$^{11}$ $b$ pairs per year. 
The challenge in a hadron collider experiments is not producing the $b$ pairs,
but rather triggering on them, selecting them and getting rid of the 
background. CDF has proven to be able to do all this with their recent
determination~\cite{cdf-2b98}
\begin{equation}
\sin 2\beta = 0.79^{+0.41}_{-0.44}
\end{equation}
using the $J/\Psi$ channel. 

Both CDF and D0 are undergoing substantial upgrades~\cite{cdf-tdr,d0-tdr}.
Among other 
improvements, CDF is getting a new, longer, vertex detector, that will 
provide 3D coordinates in a larger acceptance region. A new time-of-flight
system will allow kaon identification at low momentum, and can, therefore, be
used for kaon tagging: using the kaon charge to decide on the flavor of
the decaying $b$ hadron. More importantly, 
a new pipelined, deadtimeless trigger system will allow
purely hadronic events (no leptons) to be triggered efficiently. This will
vastly increase the physics capabilities of CDF, as it will be shown later. 
The new all-hadron trigger requires two high transverse momentum tracks 
at level 1 and, using vertex detector information, finds their vertex at
level 2 and requires the decay length to be positive.

D0 upgrades are not less important. The detector will have for the first time
a solenoidal coil, with a 2~Tesla field that, together, with the new
scintillating-fiber central tracking system, will provide precise momentum
measurements for all charged tracks. Furthermore, a new four-layer vertex
detector will also be added. These changes should allow D0 to reconstruct and
tag B decays using standard displaced vertes techniques.

A comparison between the $b$-physics capabilities of CDF and D0 shows a clear
advantadge for the former, due, in particular, to its unique 
all-hadron trigger.
In order to trigger on hadronic B decays, D0 has to rely on 
semileptonic decays of the opposite-side B hadron, therefore paying the price
of the 10\%
semileptonic branching ratio.
\subsection{$B_s$ oscillations}
As mentioned in the introduction, the measurement of the mass difference of
the two $B_s$ sates, $\Delta m_s$,
is needed in order to determine the length of
the side of the standard CKM triangle opposite the angle $\gamma$. The mass 
difference is obtained from the difference between mixed and unmixed $B_s$
decays as a function of proper time:
\begin{equation}
\frac{N_{unmixed}(t) - N_{mixed}(t)}{N_{unmixed}(t) + N_{mixed}(t)}
= D\cdot \cos\left(\Delta m_s t\right)  \ ,
\end{equation}
where $D$ is the so-called dilution factor, explained later. In order
to be able to perform the measurement one needs to determine:
\begin{itemize}
\item The proper decay time. Hence good decay length resolution is needed.
\item The $B_s$ flavor at production time (tagging). The dilution factor
$D$ in the previous
equation is $D=1-2\,p_{mistag}$, where $p_{mistag}$ 
is the probability of getting the flavor wrong.
\item The $B_s$ flavor at decay time, using, for instance, flavor-specific 
decays.
\end{itemize}
CDF has studied the flavor-specific decay channel $B^0_s\to D^-_s\pi^+$
or $B^0_s\to D^-_s\pi^+\pi^-\pi^+$ with $D^-_s\to \phi\pi^-, K^{*0}K^-$. 
As one can see, there are
no leptons and, therefore, the purely hadronic trigger is
mandatory. CDF expects about 20000 selected events in the first two years of
operation
with a signal-to-background
ratio between 1/2 and 2~\cite{cdf-kroll}. 
The significance of the measurement of $\Delta m_s$ 
(in number of standard deviations) or of
the related, dimensionless parameter $x_s$ introduced above,
is given by
\begin{equation}
Sig(x_s) = \sqrt{\frac{N\epsilon_{tag} D^2}{2}}
\exp{\left(-\left(x_s\sigma_t/\tau\right)^2/2\right)}
\sqrt{\frac{S}{S+B}}  \ ,
\end{equation}
where $N$ is the number of selected events, $\epsilon_{tag}$ is the tagging
efficiency, $D$ is the dilution factor, $\sigma_t$ is the proper time 
resolution, $\tau$ is the $B_s$ lifetime
and $S$ and $B$ are the number of signal and background 
events selected, respectively.
Looking at the formula, one sees that
the ``quality factor'' $\epsilon_{tag} D^2$ gives the effective tagging
efficiency and that good proper time resolution is, clearly, crucial. 
No matter how many events one can collect, the reach in $\Delta m_s$ is
going to be limited by the time resolution
to something of order ${\cal O}\left((1-3)/\sigma_t\right)$. CDF expects a
time 
resolution around 50~fs and a quality factor $\epsilon_{tag} D^2 = 0.113$, 
which translate into a sensitivity to $x_s < 63$ at the five standard deviation
level~\cite{cdf-kroll}.
%(fig.~3).
% 
%\begin{figure}[t]
% \vspace{9.0cm}
%\special{psfile=dafne99profig.ps voffset=-220 hoffset=-30
%         hscale=80 vscale=80 angle=0}
% \caption{\it
%      The standard CP violation triangle.
%    \label{fig3} }
%\end{figure}
%
This goes well beyond the current LEP limit of 21.8 and the
expected range in the Standard Model, from 20 to 30. In summary, CDF should be
able to determine $x_s$ quite precisely unless it is much larger than expected.
\subsection{Measurement of $\sin 2\beta$}
The standard prodecure to determine the angle $\beta$ uses the time integrated
asymmetry in the $B^0/\overline{B}^0$ decays to $J/\Psi K_s$:
\begin{eqnarray}
A_{CP}  &=& \frac{N(B^0\to J/\Psi K_s) - N(\overline{B}^0\to J/\Psi K_s)}
{N(B^0\to J/\Psi K_s) + N(\overline{B}^0\to J/\Psi K_s)}  \nonumber \\ 
        &=& -D_{mix}\sin 2\beta \ , \ \ \mbox{\rm with} \nonumber \\
D_{mix} &=& x_d/\left(1+x_d^2\right) \simeq 0.47 \ .
\end{eqnarray}
Here $D_{mix}$ is a dilution factor due to $B^0_d-\overline{B}^0_d$ mixing. 
Since
$x_s$ is much larger than $x_d$ this dilution factor is very small in the
$B^0_s$ asymmetries, so that time-dependent measurements are needed.

The experimentally observed asymmetry is further diluted by mistagging and
background:
\begin{eqnarray}
A_{obs} &=& D_{tag}\cdot D_{bgd}\cdot A_{CP} \nonumber \\
D_{tag} &=& 1 - 2\, p_{mistag}\ , \ \ \ D_{bgd} = \sqrt{S/\left(S+B\right)} \ ,
\end{eqnarray}
so that the final uncertainty on $\sin 2\beta$ can be written as
\begin{equation}
\delta\left(\sin 2\beta\right) = 
\frac{1}{D_{mix}D_{tag}}\frac{1}{\sqrt{\epsilon_{tag}N}}\sqrt{\frac{S+B}{S}}\ .
\end{equation}

CDF believes they can obtain a total $\epsilon_{tag}D^2_{tag}$ around 9.1~\%,
including the kaon tagging using the proposed time-of-flight system. With the
statistics available in the first two years of data-taking, that would 
imply $\delta(\sin 2\beta) \simeq 0.07$~\cite{cdf-kroll}. On the other hand, 
total $\epsilon_{tag}D^2_{tag}$ for D0 is expected to be around 5.0~\%,
implying $\delta(\sin 2\beta) \simeq 0.15$, using only the muon decay of 
the $J/\Psi$~\cite{d0-beta}. 
Analyses are underway to quantify the precision attainable in
the electron channel.
\subsection{Other CP Angles}
Determination of CP angles other than $\beta$ can also be attempted by CDF. It
will be more difficult for D0 because of their lack of 
a fully hadron trigger. A quick summary of CDF capabilities follows:
\begin{itemize}
\item $\sin 2\alpha$ could, in principle, be obtained from 
the $B^0-\overline{B}^0$ asymmetry in the $\pi^+\pi^-$ decay. 
However, the large
penguin contribution, with different phase, my preclude the effective
extraction of $\alpha$ from this channel. Some other methods will be described 
in section~3. In any case, CDF can have a quality factor around 9.1\%
for this channel, leading to a precision in the determination of
the asymmetry of $\delta A(\pi^+\pi^-)\simeq 0.09$~\cite{cdf-kroll}.
\item There are several ways of studying the angle $\gamma$. CDF has
explored, for instance, the use of the asymmetries in the 
decays $B^0_s/\overline{B}^0_s \to D^\pm_s K^\mp$ 
and $B^\pm\to K^\pm D^0_{CP}$,
where $D^0_{CP}$ is the $CP=1$ eigenstate of the $D^0$ system. For the moment,
however, there are no quantitative estimates of the possible $\sin 2\gamma$
reach.
\item Finally, the asymmetry in the 
decays $B^0_s/\overline{B}^0_s \to J/\Psi\phi$,
the ``golden decay'' equivalent in the $B_s$ sector, can be determined to
about 10\% 
by CDF, depending on the value of $x_s$~\cite{cdf-kroll}. 
%(fig.~4). 
This asymmetry provides a
direct determination of $\delta\gamma$.
% 
%\begin{figure}[t]
% \vspace{9.0cm}
%\special{psfile=dafne99profig.ps voffset=-220 hoffset=-30
%         hscale=80 vscale=80 angle=0}
% \caption{\it
%      The standard CP violation triangle.
%    \label{fig4} }
%\end{figure}
%
They expect about 6000 events 
with $\epsilon_{tag}D_{tag}\simeq 9.7\%$.
While only an asymmetry of order a few percent is expected in the Standard
Model, a larger value, if found, would be a sign of new physics. It should be
noted that since in this channel the triggering is based on the leptonic decays
of the $J/\Psi$, D0 should be able to do a similar job.
%although no numbers have been put forward up to now.
\end{itemize}
\section{The More Distant Future: ATLAS, CMS, LHCb, BTeV}
\subsection{The situation in 2005}
In about 2005 ATLAS, CMS and LHCb will start taking data at the LHC 
accelerator at CERN. At about the same time, BTeV, if approved, will start at
the Tevatron. At that time a lot of progress will have been made by the earlier
experiments, both at hadron machines and at the electron-positron factories.
In particular, one could expect $\sin 2\beta$ to be measured with a precision
around 0.04 by a combination of BaBar, Belle, HERA-B, CDF and D0. The length 
of the side opposite $\gamma$ will be known thanks to the $B_s$ oscillations
measurements at HERA-B, CDF and D0 but it is rather unclear the amount of
information that will be gathered about $\alpha$ and $\gamma$. The outcome of
the combination of all the different measurements should be
one of the following:
\begin{itemize}
\item either a clear inconsistency with the Standard Model predictions will
be seen in the precise measurements ($\beta$ and $B_s$ mixing); 
\item or a hint of inconsistency with the less precise measurements
($\alpha$, kaon results) will appear;
\item or all measurements will look consistent with the Standard Model.
\end{itemize}

In any case, the next generation of experiments will be needed for a full 
understanding of the CP violation mechanism. That will involve more precise
measurements of the same parameters in the same channels, measuring the same 
parameters in different channels and determining for the first time some 
parameters, notably the angle $\gamma$.
\subsection{The detectors}
Around year 2005, four new hadron-machine experiments with interesting 
capabilities in $b$ physics will start taking data. Two of them are 
specifically designed for $b$ physics, 
LHCb~\cite{lhcb-tp} and BTeV~\cite{btev-loi}. Both of them cover only
the forward region, since most of the $b\bar{b}$'s produced
are in this region. BTeV actually covers also the backward region, thus 
doubling the acceptace. The larger cross section available at LHC energies 
(500~$\mu$b vs.~100$\mu$b) more
than compensates for the smaller acceptance of LHCb. Covering only the 
forward/backward regions has clear advantages:
\begin{itemize}
\item $b\bar{b}$ production peaks forward;
\item limited solid angle coverage leads to limited cost;
\item the $b$ hadrons have higher momentum which leads to easier vertex finding
and improved decay time resolution;
\item open geometry allows for easy installation and maintenance.
\end{itemize}
These advantadges 
compensate for the higher minimum bias background, occupancies and 
radiation dose that have to be overcome at low angles.

The main advantadges that LHCb and BTeV have over the all-purpose detectors, 
ATLAS and CMS,
are in particle identification and, especially, in the triggering
capabilities. LHCb and BTeV have RICH systems that allow $\pi/K$ separation in
a large momentun range ($1<p<150$ for LHCb, $3<p<70$ GeV for BTeV). This is
mandatory to separate signal from background in the $B\to \pi\pi,KK, K\pi$
channels and adds the possibility of using kaon
tagging, as mentioned in chapter~2 for CDF.

Both LHCb and BTeV can trigger efficiently in all-hadron events, thanks to
their vertex triggers. In contrast, ATLAS and CMS need a high-$p_T$ lepton
to start the triggering sequence. This considerably reduces the
$b$-physics capabilities of the two multipurpose detectors.

%Figure~5 shows a top view drawing of the LHCb detector. The large dipole
%magnet, RICH system and electron and hadron calorimeters are particualrly
%prominent.
%% 
%\begin{figure}[t]
% \vspace{9.0cm}
%\special{psfile=dafne99profig.ps voffset=-220 hoffset=-30
%         hscale=80 vscale=80 angle=0}
% \caption{\it
%      The LHCb detector.
%    \label{fig5} }
%\end{figure}
%
%
\subsection{Physics Reach}
\subsubsection{\it $\sin 2\beta$}
The angle $\beta$ can be measured very precisely by the second generation $b$
experiments using the same ``golden'' decay used by 
CDF and D0: $B^0/\overline{B}^0\to J/\Psi K_S$. Since the $J/\Psi$ is only
reconstructed through its lepton-pair decays, the triggering and particle-id
advantadges of LHCb and BTeV are of no importance and the kind of
precision that can be achieved by the four experiments is very similar, ranging
from $\Delta\sin 2 \beta = \pm 0.015$ for 
LHCb to $\pm 0.025$ for CMS~\cite{2005-beta}, with
ATLAS and BTeV somewhere in between.
\subsubsection{\it $x_s$ reach}
The $B_s$ mixing analysis is much improved by the availability of a pure hadron
trigger and, therefore, the differences between the reach of the experiments 
are large. Whereas ATLAS and CMS could only
reach $x_s$ values around 50~\cite{2005-beta}, lower
than what can be explored already at CDF with their hadron trigger, the
dedicated experiments
can go up to 60 (BTeV) or 75 (LHCb)~\cite{2005-beta}. It should be said, 
however, that 
Standard Model expectations for $x_s$ are in the range 20--30, easily 
accessible by the first generation experiments.
%or even by HERA-B at DESY.

%Figure~6 shows a simulation by BTeV of the proper time distribution 
%in the $B^0_s\to D_s^-\pi^+, \overline{B}^0_s\to D_s^+\pi^-$ channel. A
%value $x_s=40$ has been assumed, together with a proper time 
%resolution of 50~fs. The rapid oscillation can be clearly seen for both mixed
%and unmixed decays.
%% 
%\begin{figure}[t]
% \vspace{9.0cm}
%\special{psfile=dafne99profig.ps voffset=-220 hoffset=-30
%         hscale=80 vscale=80 angle=0}
% \caption{\it
%    $B_s$ mixing in the $D_s\pi$ channel as simulated by the BTeV experiment.
%    \label{fig6} }
%\end{figure}
%
\subsubsection{\it $\sin 2\alpha$}
The asymmetry in the $B^0_d\to \pi^+\pi^-$ channel has already been mentioned
in section~2 as a possibility for measuring $\alpha$, although, at present,
its usefulness is not clear due to the large penguin 
contribution with different phase. In any event,
LHCb and BTeV can measure the asymmetry in this all-hadron channel 
with a $\pm 0.025$ precision~\cite{2005-beta}. 

More promising, if more involved, seems to be 
the $B^0_d/\overline{B}^0_d\to\pi^+\pi^-\pi^0$
channel, which proceeds through $\rho$ intermediate states. Here, all 
amplitudes can be determined and the remaining ambiguities can be resolved
using the intereference regions in the Dalitz plot~\cite{Dalitz}.
The analysis seems feasible, but no experiment has yet made public the
precision it could achieve in $\alpha$.
\subsubsection{\it $\sin 2\gamma$}
There are several ways of getting to the angle $\gamma$, all of them rather
challenging. For example, it can be obtained from the simultaneous measurement
of six time-integrated decay rates in 
the $B^0_d\to D^0 K^{*0}$ channel:
$B^0_d\to D^0 K^{*0},B^0_d\to \overline{D}^0 K^{*0},
B^0_d\to D^0_{CP} K^{*0},
\overline{B}^0_d\to \overline{D}^0 \overline{K}^{*0},
\overline{B}^0_d\to D^0 \overline{K}^{*0},
\overline{B}^0_d\to D^0_{CP} \overline{K}^{*0}$, 
with $D^0$ decaying to $K^-\pi^+$, $\overline{D}^0$ to $K^+\pi^-$,
and $D^0_{CP}$ to either $K^+K^-$ or $\pi^+\pi^-$. The final state, 
therefore, consists in all cases in four charged kaons or pions resonating in 
different masses. It is clear that proper pion/kaon separation is crucial.
%as can be seen from fig.~7, showing the $B^0_d$ reconstructed by the LHCb
%simulation in this channel with and without the use of the RICH detector. 
%% 
%\begin{figure}[t]
% \vspace{9.0cm}
%\special{psfile=dafne99profig.ps voffset=-220 hoffset=-30
%         hscale=80 vscale=80 angle=0}
% \caption{\it
%     Reconstruction of the $B^0_d\to D^0 K^{*0}$ decay in LHCb.
%    \label{fig7} }
%\end{figure}
%
LHCb claims a precision in $\gamma$
around 10$^{\rm o}$ from this channel~\cite{lhcb-tp}.

BTeV has studied a similar 
channel: $B^-\to D^0/\overline{D}^0 K^-$, where $D^0$ 
and $\overline{D}^0$ go to the same final state, 
and the corresponding $B^+$ decays.
Using now nine time-integrated decays, the precision in $\gamma$ is found to be
about 13$^{\rm o}$~\cite{btev-gamma}.  
\subsubsection{\it Other CP angles}
Apart from $\alpha$, $\beta$ and $\gamma$,
other CP angles are also accessible to the second generation CP experiments:
\begin{itemize}
\item The small angle $\delta\gamma$ introduced in section~1 can be measured
precisely using the $B^0_s/\overline{B}^0_s \to J/\Psi\phi$ asymmetry 
as already
mentioned above. Precisions around $0.5-0.9^{\rm o}$ seem possible, even for
ATLAS and CMS~\cite{2005-beta}, 
because of the leptonic decays of the $J/\Psi$.
\item The combination $\gamma -2\delta\gamma$ can be determined with a
precision around 10$^{\rm o}$~\cite{2005-beta}
by both BTeV and LHCb through the 
decay $B^0_s\to D^-_s K^+$ and its charge conjugated. Since $\delta\gamma$ is
expected to be small in the Standard Model, this can be viewed as yet another
way of getting $\gamma$, only in this case through $B_s$ decays, and, hence,
a very interesting check.
\item LHCb has also studied the extraction 
of $2\beta+\gamma = \pi+\beta-\alpha$ 
from $B^0\to D^{*+}\pi^-$. The precision 
that can be obtained is around 9$^{\rm o}$~\cite{2005-beta}.
Again, since $\beta$ will be well known, this method could be used to 
obtain $\gamma$ or $\alpha$.
\end{itemize}
The ability of measuring a single angle with different processes can be very 
useful in checking whether the Standard Model by itself is able to explain CP
violation.
\section{Summary}
This talk has tried to convey the message that hadron machines have a 
very comprehensive program for understanding the origin of CP violation in the
period 2001--2015 or so, at the Tevatron and LHC.

In a first phase, starting in 2001,
CDF and, to a lesser extend, D0 will be able to 
measure $\sin 2\beta$ to about 0.07 and study $B_s$ mixing for values of
the mixing parameter $x_s$ up to 63.

In a second phase, from 2005 on, LHCb, BTeV, if approved, plus ATLAS and CMS
will improve on $\beta$ and $x_s$
and will be able to determine both $\alpha$ 
and $\gamma$ to about 10$^{\rm o}$, thus putting strong constraints on the
Standard Model ability to explain all the phenomenology of CP violation in 
the $b$ sector.
\section{Acknowledgements}
It is a pleasure to thank Giorgio Capon and the rest of the organizers of the
workshop for the kind 
invitation to give this talk and for their patience while
waiting for me to finish writing this manuscript.
\end{document}